# Sub-µW Battery-Less and Oscillator-Less WiFi Backscattering Transmitter Reusing RF Signal for Harvesting, Communications and Motion Detection

Marco Privitera, *Member, IEEE,* Andrea Ballo, *Member, IEEE,* Karim Ali Ahmed, *Member, IEEE,*
Alfio Dario Grasso, *Senior, IEEE,* Massimo Alioto, *Fellow, IEEE*

***Abstract*— In this paper, a sub-µW power 802.11b backscattering transmitter is presented to enable reuse of the same incident wave for three purposes: RF harvesting, backscattering communications and position/motion sensing. The removal of the battery and any off-chip motion sensor (e.g., MEMS) enables unprecedented level of miniaturization and ubiquity, unrestricted device lifespan, low fabrication and maintenance cost. The µW power wall for WiFi transmitters is broken for the first time via local oscillator elimination, as achieved by extracting its frequency through second-order intermodulation of a two-tone incident wave. The two-tone scheme also enables a cumulative harvesting/transmission/sensing sensitivity down to $P_{min}$~ -19 dBm. Position/motion sensing is enabled by using the harvested voltage as a proxy for the Received Signal Strength (RSS), allowing to sense the chip location with respect to the tone generator(s) shared across tags in indoor neighborhoods.**

***Index Terms*— Radiofrequency, backscattering, positioning system, motion detection, WiFi, battery-less, green system s**

## I. INTRODUCTION

RELENTLESS advancements in circuits for battery-less silicon systems have recently enabled new levels of sensor node miniaturization, cost and device lifespan, removing their limitations imposed by batteries [1], [2]. Being the available RF harvested power in the deep sub-µW/mm$^2$ range, peak power budgets are restricted to the 1 µW range or less at mm-scale form factors, requiring aggressive system power reductions [1]. At the battery-less system level, wireless transmission (e.g., WiFi) generally sets the limit to further system power reductions in view of their dominant peak power [3], [4].
The mW-range power consumption of state-of-the-art conventional WiFi transmitters [5]-[7] was recently reduced to tens of µWs through backscattering architectures. Indeed, the latter suppress GHz-range PLL and power amplifier by properly reflecting an incident wave from a tone generator [8]. Complete PLL elimination has recently reduced WiFi transmitter power to µWs, thanks to event-driven temperature-compensated ring oscillator-based local oscillators [4], [9]-[11]. The sizeable local oscillator power sets an impediment to further power reductions due to the phase noise requirements imposed by the wireless standard [3], [12], [13], and its removal hence requires further innovation. Furthermore, the adoption of the 802.11b WiFi standard is preferred to BLE, LTE and 5G since these standards do not have widely distributed infrastructure in most of the indoor spaces (e.g., home, offices, gyms) which makes rapid low-cost deployment quite difficult [3].

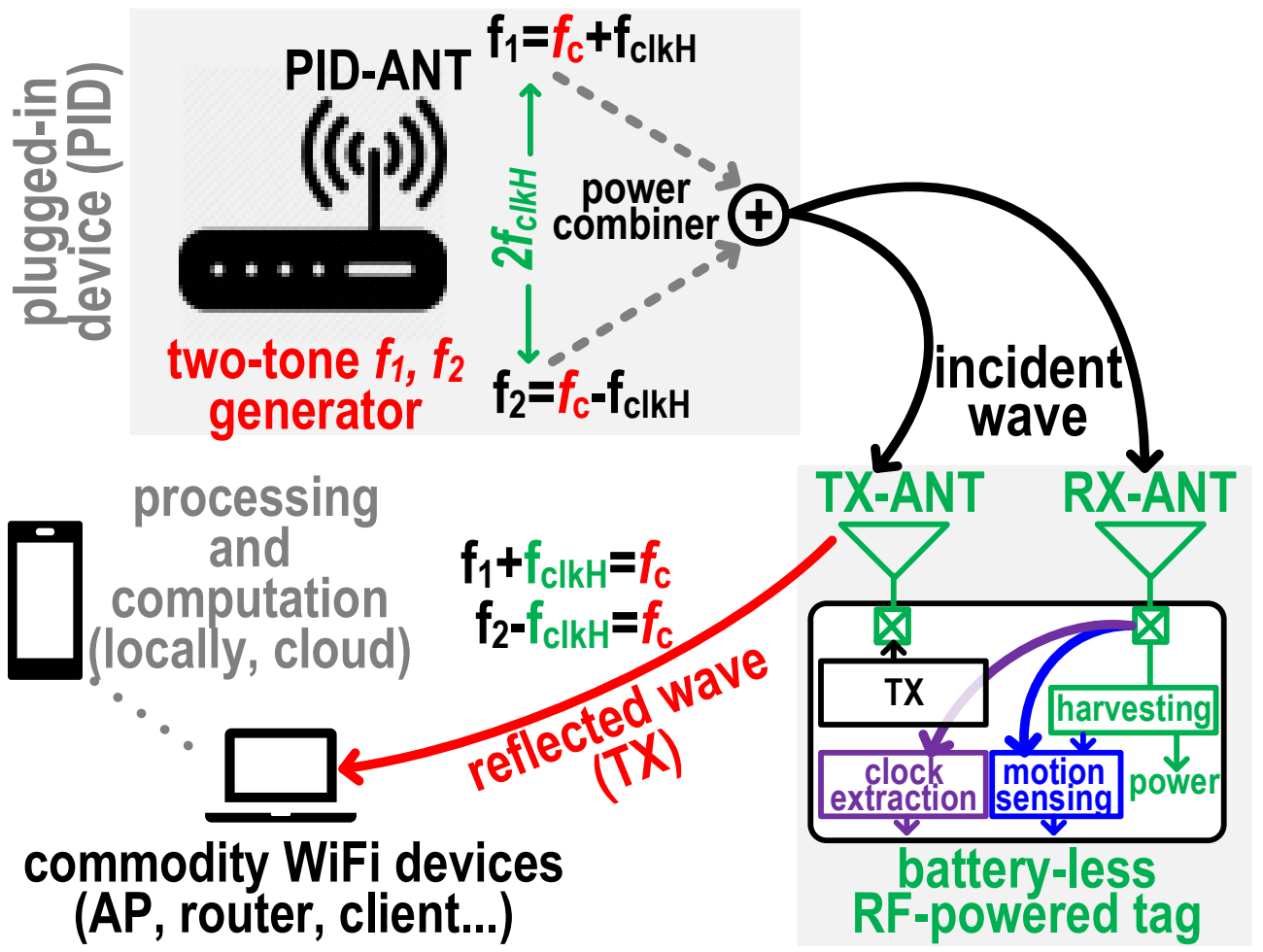

Fig. 1. Two-tone passive WiFi enables direct clock extraction from the RF signal sent by the tone generator (PID, plugged-in device), simplifies harvesting, and enables communications with commodity WiFi devices.

The goal of this work is to enable a new class of WiFi backscattering transmitters with power in the sub-µW range by removing the local oscillator power, while simultaneously and for the first time in the literature reusing the incoming RF signal for backscattering communications, harvesting, and sensor-less position/motion sensing.

Local oscillator and PLL power are eliminated via second order intermodulation (*IM2*) clock extraction from a two-tone incident wave (Fig. 1), which is also re-used to achieve -19 dBm overall communications/harvesting/sensing sensitivity.

Manuscript received May 7, 2025. This work was supported by the Singapore MOE (grant MOE-T2EP50123-0022), by TSMC for testchip fabrication, and in part by European Union's Horizon Europe research and innovation program under the Marie Skłodowska-Curie grant agreement no. 101086359. At the time of this work, Marco Privitera, Andrea Ballo, Karim Ali Ahmed and Massimo Alioto are with the ECE department at the National University of Singapore (e-mail: massimo.alioto@nus.edu.sg), Alfio Dario Grasso is with the DIEEI Department at the University of Catania, which Marco Privitera and Andrea Ballo are affiliated to.

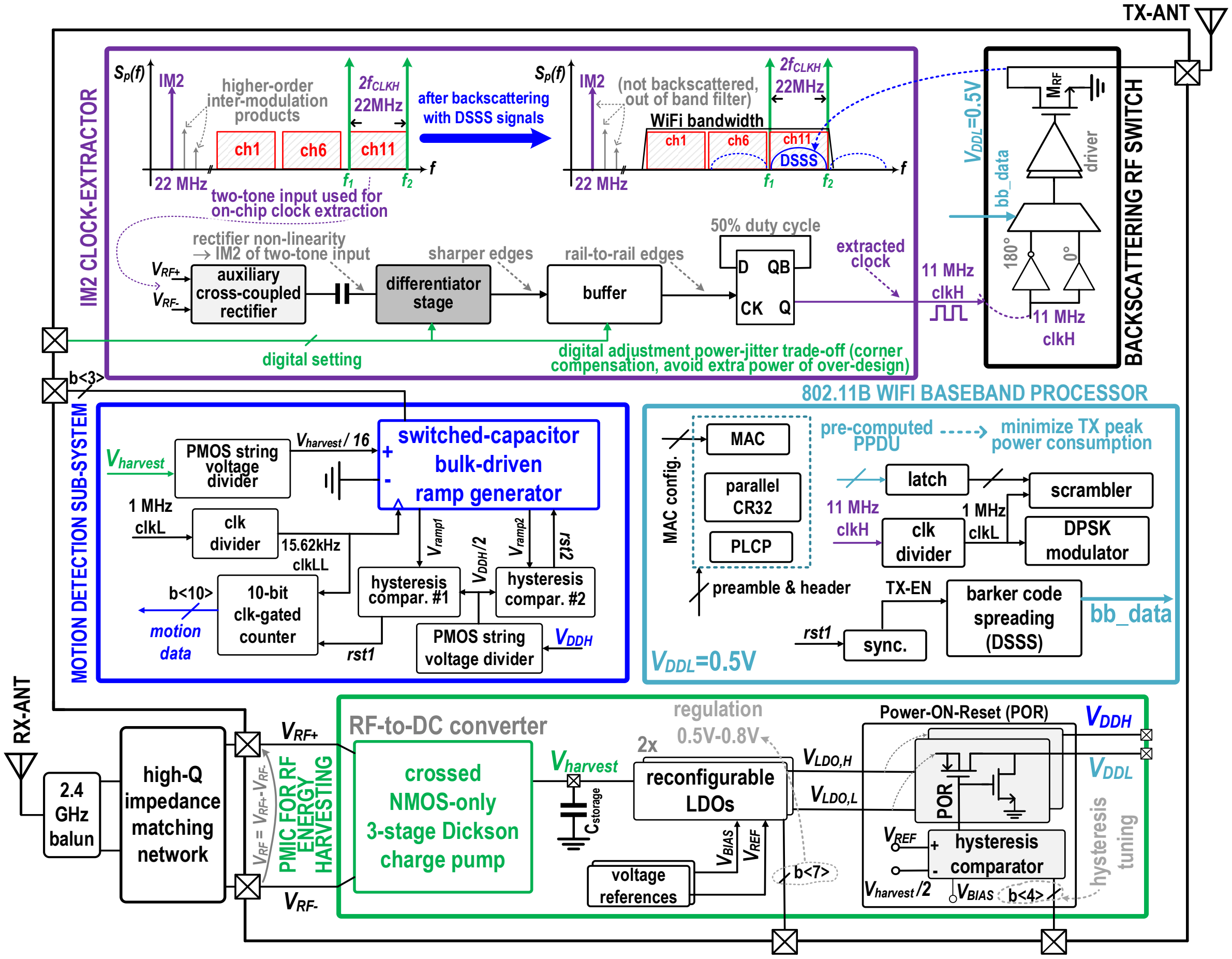

Fig. 2. System architecture including clock extraction from incoming two-tone RF incident signal (top-left), backscattered transmission via RF switch impedance modulation (top-right), sensor-less motion detection based on harvested voltage set by distance from PID (mid-left), 802.11b baseband processor for packet assembly (mid-right), and harvesting (bottom) embedding a novel crossed-NMOS Dickson charge pump.

Compared to previous arts adopting intermodulation-based clock-extraction circuitry, the proposed solution does not require any filters, any noisy and power-hungry injection-locked ring-oscillators and any limiting amplifiers [10], [14]-[18]. Thanks to its simplicity clock-extraction circuit allows, hence, to aggressively reduce the power consumption down to few hundreds (or less) nanoWatt. In addition, the circuit is digitally adjustable to always meet the jitter/phase noise requirements of the adopted 802.11b WiFi standard against process corners variations.

The two-tone incident wave is also exploited, for the first-time, for WiFi backscattering communication. Compared to single-tone (i.e., from PID) backscattering radios [4], [8], [11], the adoption of two input tones gives an improvement of more than 3 dB on the signal-to-noise ratio (SNR) at the receiver-end and, hence, an increased working tag-to-receiver distance range (up to 25 m) within standard compliant BER $\leqslant 10^{-5}$.

Inexpensive rectification and amplification of the RSS voltage is obtained via a novel and efficient crossed-NMOS Dickson charge pump exploited for the RF energy harvesting operation and fully battery-less. Thanks to this topology that mixes the benefits of cross-coupled charge pumps (e.g., low-input sensitivity) with Dickson topologies (e.g., higher output current and higher efficiency levels), the best-in-class 1V-sensitivity is achieved with an input power down to -20.3 dBm. The 43% peak power conversion efficiency (PCE) is reached with the lowest input power compared to similar works dealing with the 2.4 GHz-band RF-to-DC converters [19]-[23]. The harvested voltage is used as a proxy for Received Signal Strength (RSS) to quantify the chip location with respect to the two-tone generator(s). The achieved resolution, also under interference signals, goes down to 16 cm that is 3X better than previous received signal strength indicator (RSSI) and channel state information (CSI) position detection systems [24]-[26].

This work is structured as follows. The system architecture and main sub-blocks are introduced and analyzed in Section II. Measurement results are discussed and compared with previous works in Section III. Finally, conclusions are drawn in Section 0V.

## II. System Architecture and Main Sub-blocks

The proposed system architecture in Fig. 2 comprises the RF harvesting and power management, the sensor-less motion detection, and the 802.11b WiFi backscattering transmitter comprising the clock extractor, the baseband processor and the RF switch backscattering the incident two-tone wave. The latter is generated by the plugged-in device (PID, serving as tone generator as in [4], [11]) as shown in Fig. 1, which is shared across all sensor nodes within range. The PID also handles their communications with surrounding access points, routers and clients as in [4], [13], [14]. The incident wave from the PID has two tones at frequency $f_1$ and $f_2$, whose difference is set to twice the desired system clock and local oscillator frequency of 11 MHz, as necessary for WiFi backscattering [10]-[14] (see Section IIB).

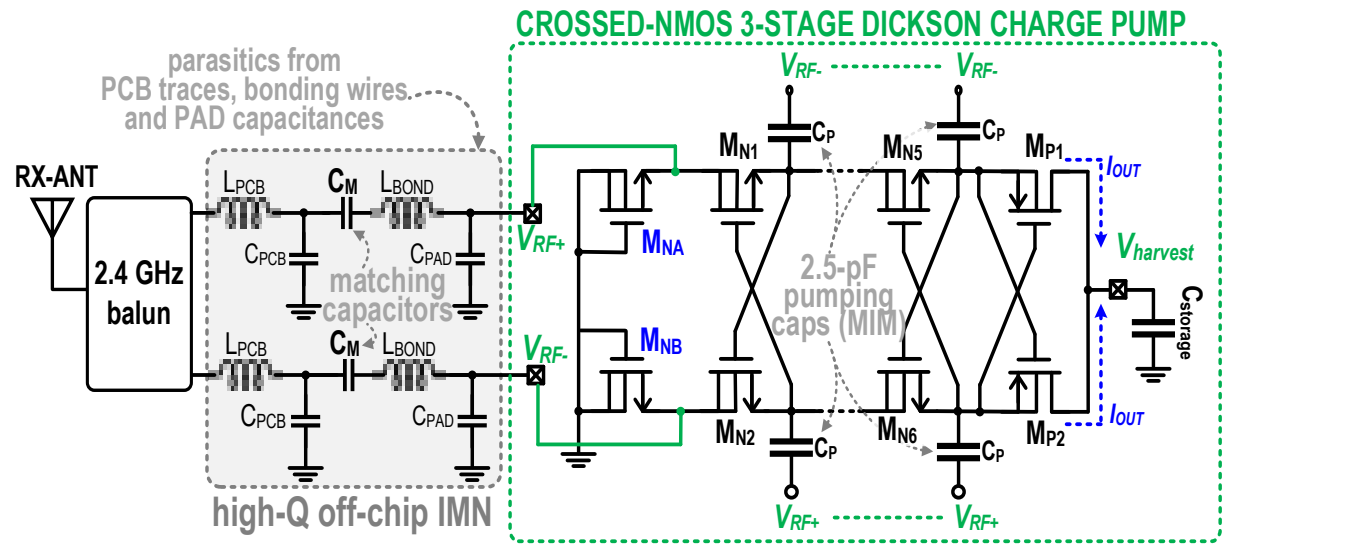

Fig. 3. Circuit details on proposed crossed-NMOS Dickson charge pump for efficient RF-to-DC conversion with sensitivity target down to ~-20 dBm.

The adoption of two antennas in Fig. 1 allows simultaneous RF harvesting, motion/position sensing and clock extraction while transmitting, as opposed to all prior backscattering tags that used extra antennas for beam forming with higher effective antenna gain [27], [28], two/three antennas for wake-up/harvesting/backscattering [13], [29], additional off-chip clock [9], [14]. At the same time, simultaneous clock extraction and transmission eliminates the need for any on-chip power-hungry clock, such as the off-chip 4 MHz clock in [14] (whose power is already ~1 μW in the adopted technology, based on a temperature-compensated ring oscillator shown in [4]). It is also worth noting that the adoption of an uncommonly low (for the BLE standard) 4 MHz clock in [14] leads to direct image interferences backscattered in other channels that is not allowed by the BLE protocol [11]. In turn, the proposed clock extractor removes the power floor imposed by the local oscillator and hence reduce power below the μW level (see Section IIIB).

*A. RF-Harvesting and PMIC*

From the Passive WiFi scheme in Fig. 1([4], [8], [11]) the PID sends two-tone RF power at FCC-compliant level (<30 dBm EIRP) to the surrounding sensor nodes at the 2.4 GHz band. The RF power reaches the considered tag, which captures it through the RX-ANT antenna in Fig. 1. As detailed in Fig. 2, the received single-ended voltage is converted to differential via a 2.4-GHz balun. A series capacitive impedance matching network is inserted as in Fig. 3 to leverage the existing inductive contributions from PCB and bonding leads to provide 14-dB of passive voltage gain of $Q/2$ [30] ($Q$ is the overall quality factor for the matching network). This extends system operation to lower incident RF power level (targeted down to ~-20 dBm).

The RF-to-DC conversion of the received differential input voltage $V_{RF} = V_{RF+} - V_{RF-}$ in Fig. 2 for RF harvesting is carried out by the proposed novel energy-efficient crossed-NMOS 3-stage Dickson charge pump in Fig. 3. The latter consists of negative voltage rectifiers $M_{NA}$-$M_{NB}$ followed by three cascaded crossed NMOS-only charge transfer switches (CTS). The rectifiers define the local ground and set the input impedance $R_{REC}$ of the RF harvester, whereas the CTS switches generate the boosted voltage $V_{harvest}$≥ 0.6 V for the intended range as system supply.

Compared to a conventional cross-coupled NMOS/PMOS charge pump [19]-[21], [18] the proposed topology in Fig. 3 improves end-to-end efficiency in the RF-to-DC conversion, especially at low input power levels. Indeed, it reduces cross-conduction and leakage losses thanks to aggressive reverse gate biasing in of $M_{N1}$-$M_{N6}$ with $V_{GS1\ldots6} = -2V_{RF}$ in the OFF state, as opposed to $-V_{RF}$ in conventional Dickson charge pumps. Also, the removal of the distributed series PMOS transistors reduces inter-stage parasitic capacitance, and suppresses the extra n-well spacing area and wire parasitics conventionally needed to prevent proximity effects. This enables more favorable balance with charge pump resistance, compared to prior art [19]-[21]. This also allows more favorable tradeoff among wake-up time, output current and higher efficiency at lower input power level (see Section IIIA). From post-layout simulations, the topology in Fig. 3 improves power efficiency by more than three times over cross-coupled NMOS-PMOS (13.6%) at same area and load current.

The charge transfer switches in Fig. 3 are followed by low-$V_T$ PMOS transistors $M_{P1}$-$M_{P2}$ for one-way current provisioning. The resulting conversion efficiency and harvesting sensitivity improvement (see Section IIIA) are further enhanced by the above-discussed adoption of a two-tone incident wave, as the RF input amplitude is increased by $\sqrt{2}$ compared to a conventional single tone operating at same single tone power and hence same tag-to-PID range.

The adoption of just three stages of CTS derives from an optimization strategy. Indeed, the main counter effect of increasing the number of stages is the increase of switching-losses and, hence, first a reduction of the PCE, and then, above 6-8 stage also the output rectified voltage is reduced at a fixed load current (e.g., increase in the equivalent charge pump resistance). On the other side, the reduction of the number of stages would lead to an insufficient output voltage at a fixed input power (tag-to-PID operating distance).

The output voltage of the charge pump is regulated via two PMOS low-dropout (LDO) regulators providing the supply voltage $V_{DDH}$=0.6 V for the analog and $V_{DDL}$=0.5 V for the digital sub-system (Fig. 2). The LDO error amplifier is a single-stage folded-cascode OTA with low-voltage current mirrors and 1.5-nA bias current, resulting in an open-loop gain of ~78 dB. As knob to save peak power across process corners, the LDO output voltage is digitally selected within the 0.5-0.8 V range with a step of 50 mV via simple voltage selection in the feedback voltage divider (PMOS-diode voltage string, drawing a 600-pA current) [31].

Auxiliary voltages $V_{BIAS}$ =0.15 V and $V_{REF}$ =0.34 V supporting the LDOs in Fig. 2 are respectively provided by 6-transistor and process-trimmed 2-transistor voltage references [32], [33]. The former reduces the line sensitivity of the bias voltage given its direct connection with the unregulated supply, whereas the latter enforces temperature independence [31].

The harvested power is then routed to the system once sufficient intentional (in-range) power becomes available, as assessed by the Power-On-Reset in Fig. 2 upon the condition $V_{harvest}$>0.48 V. This is done via hysteresis comparator with 1-nA bias current for ultra-low power consumption and delay time in the order of 100 µs. Indeed, the overall PMIC time-response (e.g., rise time of the charge pump with LDO and POR including the cold-start condition) has always to be less than typical 10 ms-range WiFi latency.

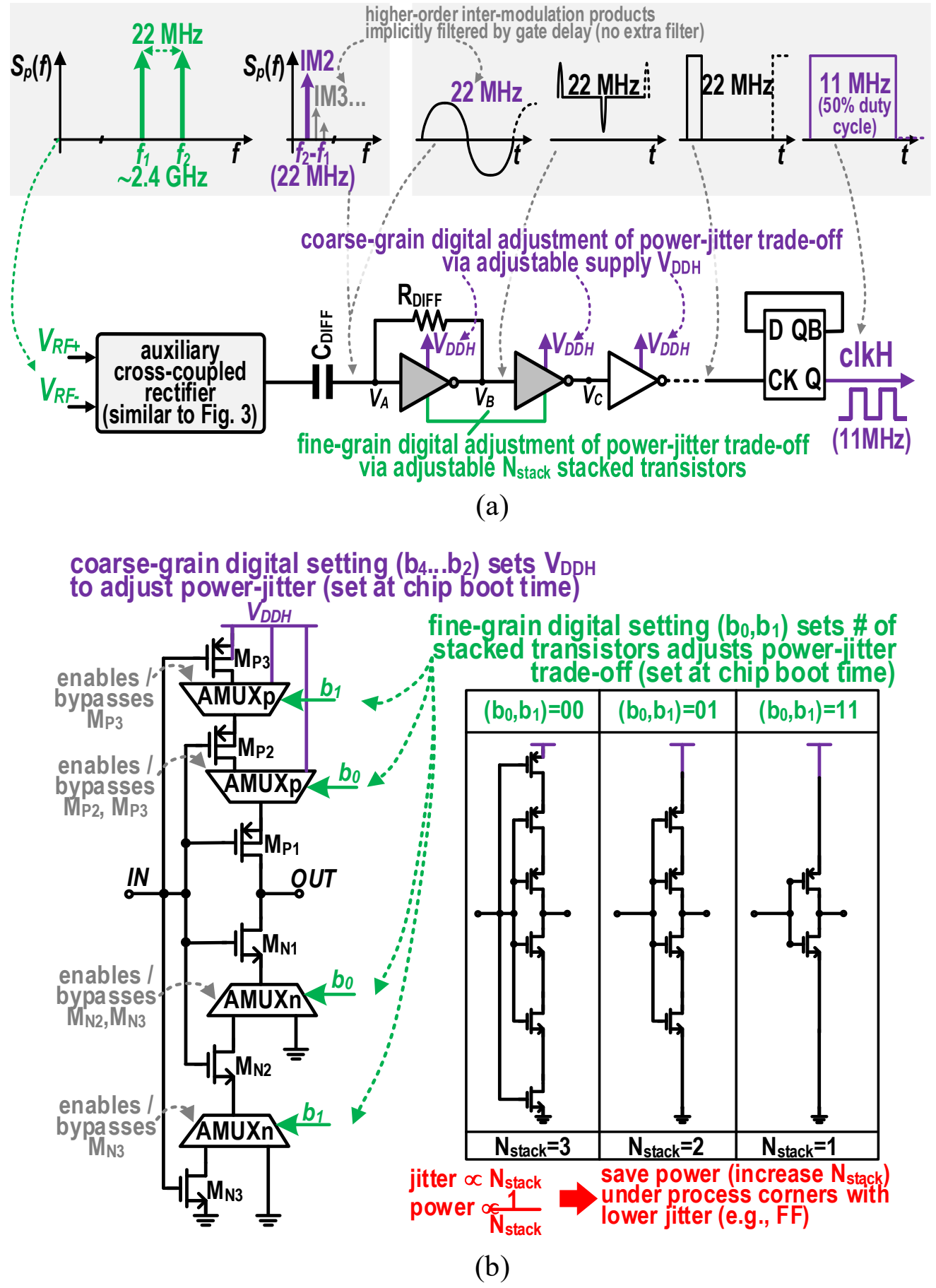

Fig. 4. a) Clock extraction based on $IM2$ inter-modulation from two-tone incident wave, b) power-jitter adjustment against power wastage at fast corners

Once the tag is addressed by the PID within range, it is powered up through the Power-On-Reset mechanism discussed about. This removes the necessity of an always-on wake-up receiver complying with the 802.11g/n/ba standard for WiFi wireless environments [34]-[38], saving both area and power.

*B. IM2-Based Clock Extraction*

The two-tone incident wave is used to extract the clock to remove the on-chip local oscillator (11 MHz for WiFi backscattering [4], [8]), whose power is constrained by jitter and stability requirements from the WiFi standard. From Fig. 4a, the clock is extracted from two tones at frequencies $f_1$ and $f_2$ through their inter-modulation products [10], [14]-[18], as generated through the non-linearity of the rectifier in the same figure. The auxiliary rectifier in Fig. 4a is a conventional cross-coupled [19] and unfiltered rectifier with minimum size to minimize its power overhead, as the DC power transfer efficiency is not a concern in the case of envelope detection in clock signal extraction. The 2nd-order inter-modulation product $IM2 = |f_1 - f_2| = 2f_{clkH}$=22 MHz is inherently isolated from the upper-order ones through adjustment of the subsequent gate delays, which effectively act as low-pass filters to eliminate conventional filtering and hence reduce power.

The near-sinusoidal $IM2$ signal in Fig. 4a is then amplified and made sharper by the subsequent reference-less inverter-based differentiator. The latter establishes the output common-mode voltage around the subsequent inverter logic threshold ~$V_{DD}/2$. Then, its voltage swing is restored rail-to-rail via the subsequent inverter gates. 2:1 frequency division restores 50% nominal duty cycle and its stability over time.

Although the jitter and stability requirements of $IM2$ are trivially met by the oscillators in the tone generator, the clock extraction circuitry in Fig. 4a can still degrade the jitter performance due to its input-referred noise. In other words, sufficient power (bias current) needs to flow through the differentiator stage and the subsequent inverter gate level restorers to keep jitter within the WiFi specifications. In turn, the power consumption of such circuitry is sensitive to the process corner, and is unnecessarily higher at faster corners (e.g., FF). To avoid the power overhead associated with across-corner margining while still meeting the jitter requirement, the jitter-power tradeoff is explicitly handled via digital adjustment of the strength of the differentiator and subsequent inverter gates as in Figs. 4a-b, which is accomplished as in Fig. 4b via:

- coarse-grain tuning of the LDO output voltage $V_{DDH}$
- fine-grain tuning of the number $N_{stack}$ of equally-sized stacked transistors in the differentiator and inverter gate stages as they both determine the bias current at the DC input value (approximately $V_{DDH}/2$).

These settings are defined at chip boot time based on the die process corner, which is identified via simple frequency comparison of an on-chip ring oscillator and the clock extracted from the two tones.

The adopted clock extraction brings further benefits at the system level. First, unintended tag operation is prevented even when an unusually high level of incoming power is received in the intended band, and unintentionally triggers system Power-On-Reset and subsequent response as in Section IIA. Indeed, even in this case the tag starts its operation only if a sufficient number of cycles is extracted from $IM2$ to leave its reset state, which in turn occurs only if there are two simultaneous tones with similar power amplitude (as required to extract $IM2$) and consistent frequency difference. As a result, unintended tag operation is very unlikely, as experimentally verified in a laboratory environment with tens of WiFi clients, smartphones, router and access point. No unintended tag wake-up was observed during repeated experiments in 30-minute windows, corresponding to a false wake-up probability better than $10^{-5}$, based on the 100-ms communication cycle of each tag.

As further benefit, the proposed clock extractor suppresses the area and the power burden of any additional active circuitry (e.g., charge domain GFSK demodulator and limiting amplifier [14]), large passive components (e.g., low-pass filter [14] and band-pass filter in [10], [15]-[1830]), noisy and power hungry injection-locked ring oscillators [10], [15]-[18], [29] in prior backscattering transmitters employing *IM2* clock extraction techniques.

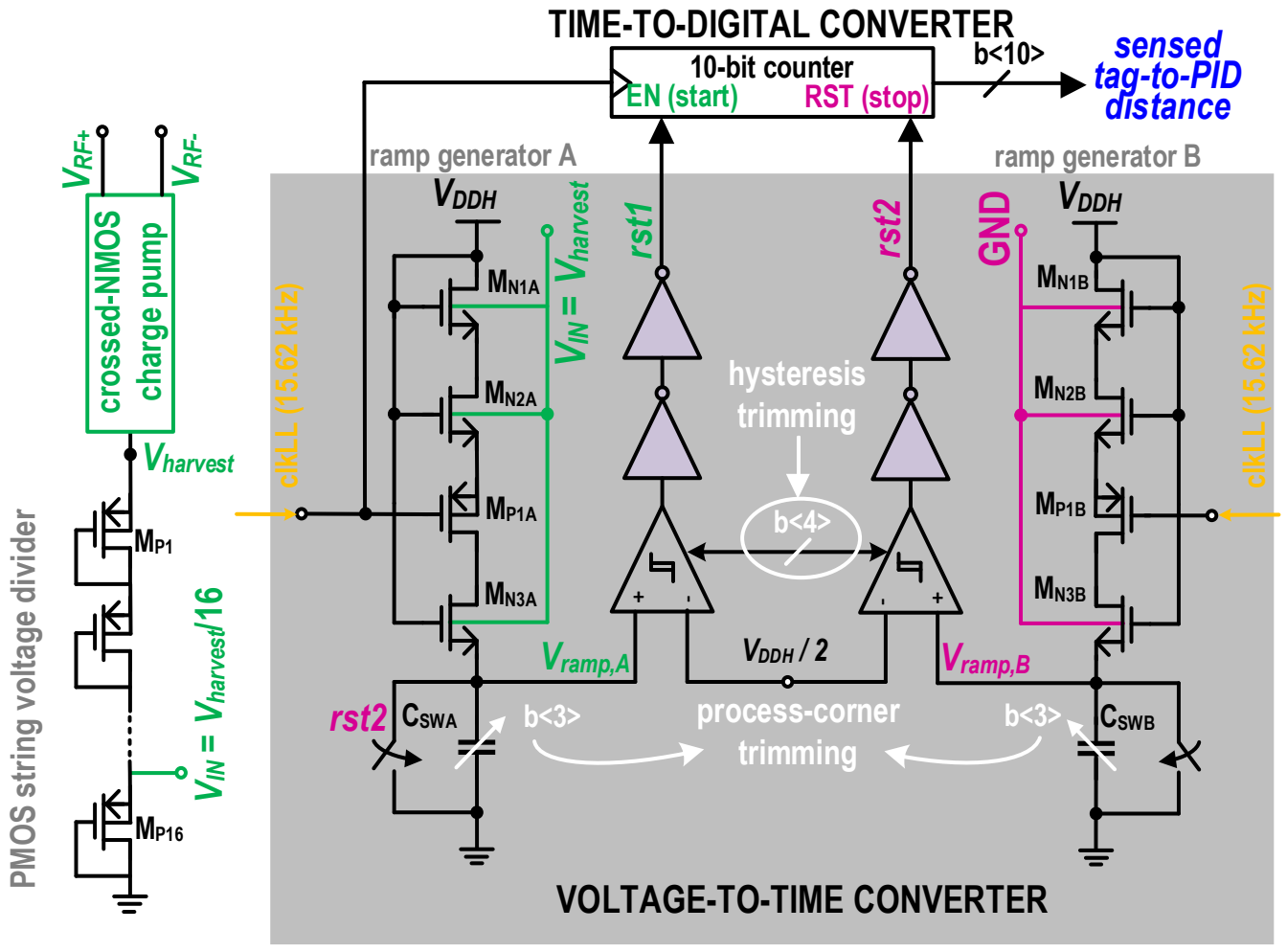

Fig. 5. Position/motion detection: voltage-to-time/time-to-digital conversion

### C. Sensor-Less Position/Motion Detection

The harvested received signal strength (RSS) [24], [25] of the incoming wave from the PID allows to perform motion/position sensing with no explicit sensor (sensor-less). To this aim, the available $V_{harvest}$ is used as a proxy for distance from the PID. This allows to check if a tag is within the range of a PID and its distance in a battery-less fashion, while supporting the many applications of conventional battery-powered tags [26] (e.g., asset tracking, personal item recovery, inventory management).

From the block diagram in Fig. 5, signal strength is inexpensively measured via differential voltage-to-time conversion of $V_{harvest}$ followed by time-to-digital conversion. Voltage-to-time conversion is carried out by two switched-capacitor ramp generators driven from their body terminal for low-voltage and supply voltage variation-resilient operation. The bulk of the first (second) ramp generator $M_{N1A}$-$M_{N3A}$ ($M_{N1B}$-$M_{N3B}$) in Fig. 5 is driven by $V_{harvest}/16$ (ground). For any practical speed of physical objects and humans in a typical living space, the tag-to-PID distance is essentially constant during a position/motion detection cycle, as the latter is in the 100-ms scale or slower (see Fig. 9b).

Hence, $V_{harvest}$ is effectively a near-DC signal whose increase determines a decrease in the threshold voltage of transistors $M_{N1A}$-$M_{N3A}$, due to their more pronounced forward body biasing. Compared to the fixed body voltage in $M_{N1B}$-$M_{N3B}$, the ramp generator driven by $V_{harvest}$ delivers a different current on its load capacitor $C_{SW}$, and hence a different ramp slope in $V_{ramp,A}$ compared to $V_{ramp,B}$ in Fig. 5. Accordingly, $V_{harvest}$ univocally determines the time window defined by the crossing of the two ramps of the $V_{DDH}/2$ value, as detected by the hysteresis comparators in Fig. 5. The series-connected NMOS-PMOS stacks MN2A-MP1A and MN2B-MP1B are inserted to reduce the on-current within each ramp generator, and hence the peak consumption.

For simplicity, only the nearly-linear range of the harvested voltage-distance is considered in the following[1]. The above time window is converted to digital via the 10-bit counter in the same figure, which counts a 15.625-kHz clock derived via frequency division from the 1-MHz clock available in the baseband processor (see Fig. 2). Hysteresis trimming suppresses process variations, whereas differential operation suppresses the impact of voltage and temperature variations.

### D. Two-tone Backscattering RF Transmission

Backscattering signal modulation for RF transmission in Fig 1. is performed by a baseband processor with sub-leakage consumption as in [4]. The processor generates the differential binary phase shift keying (DBPSK) signal that drives the RF switch to backscatter the incident two-tone signal, and ultimately generate a WiFi-compliant reflected signal. The two 0-$\pi$ phase values for DBPSK modulation are simply derived via multiplexing of the extracted clock or its complement. The multiplexed output is buffered to drive the NMOS RF switch.

Since the incoming wave contains two tones at frequencies $f_1$ and $f_2$, backscattering generates four images at frequency $f_1 \pm f_{clkH}$ and $f_2 \pm f_{clkH}$. The two images at $f_1 + f_{clkH}$ and $f_2 - f_{clkH}$ constructively align at the center frequency giving an increase of the signal power of the targeted 802.11b non-overlapped channel of $\geqslant$ 3 dB compared to single-tone backscattering (Fig. 6a,c), and hence an increase in the SNR (bit-error rate, BER) at a fixed tag-to-PID distance. It is worth noting that the other two images at frequency $f_1 - f_{clkH}$ and $f_1 + f_{clkH}$ fall in the adjacent channels as shown in Fig. 6a (similarly to [8] and [14]) and Fig. 6c.
Furthermore, the selection of $\Delta f = f_2 - f_1 = 22$ MHz allows to generate an 802.11b-compliant DSSS spectrum, while the selection of $\Delta f = f_2 - f_1 = 11$ MHz generates spectrum images overfilling the adjacent channels, not well confined in the target one selected by the router (shown in Fig. 6b) and resulting hence in a non-compliant WiFi packet transmission, since the spectral mask requirements are not fulfilled.

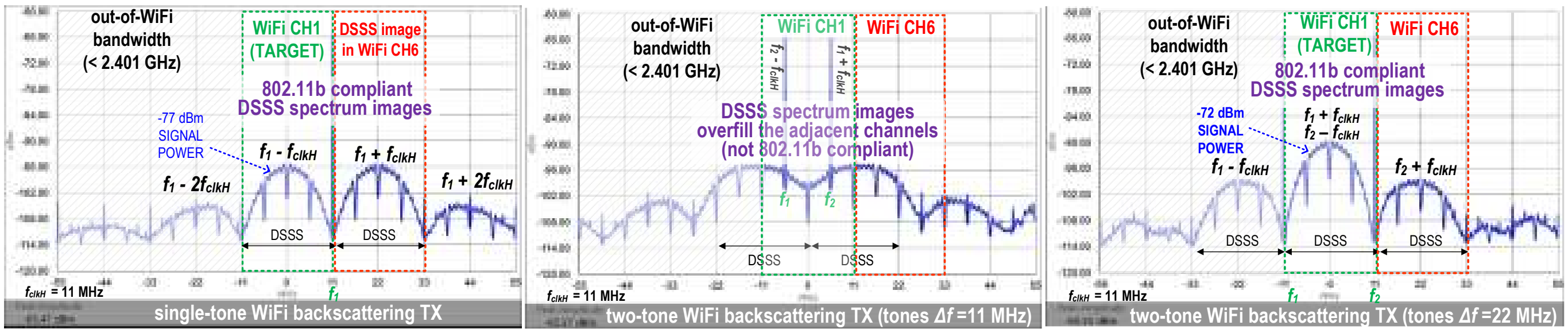

(a) (b) (c)

Fig. 6. (a) Single-tone backscattering transmission and spectrum as in [4], [8] and [11] (BLE), (b) two-tone backscattering with $\Delta f$=11 MHz between the two-incident tones and spectrum not 802.11b-compliant (images overfilling adjacent channels, spectral mask requirements not fulfilled), (c) proposed two-tone backscattering with $\Delta f$=22 MHz with 802.11b-compliant DSSS spectrum and images overlapping in target channel (CH1) for an increased SNR and BER.

[1] Extension to a wider range and non-linear dependence is straightforward, as it simply requires look-up table mapping of $V_{harvest}$ onto tag-to-PID distance.

As illustrated in Fig. 2 for non-overlapping channels, the backscattered signal occupies the intended channel (e.g., WiFi channel #11) and the one adjacent to it (e.g., channel #6), whereas its mirrored image falls outside the filter band and is hence suppressed. This requires the PID device to perform carrier sensing in the adjacent channel, before initiating a tag response in a targeted channel [4], [8]. The increased spectrum usage is not an issue in most applications, in view of the typically short and infrequent packet transmissions from tags.

The RF switch was sized to maximize the backscattered signal transmitted by the antenna TX-ANT in Fig. 2, and ultimate increase the transmission range. In detail, the backscattered signal power is well known to be proportional to $|\Gamma_1^* * -\Gamma_0^*|^2$, where $\Gamma_1^*$ and $\Gamma_0^*$ are the complex conjugate of the reflection coefficients at TX-ANT associated with the two binary symbols to be transmitted [8]. To this aim, $\Gamma_1^*$ and $\Gamma_0^*$ are set by the RF switch impedance in the two states such that the RF switch in Fig. 2 is 50 Ω for symbol 1 when turned ON, and near open-circuit when OFF for symbol 0.

Regarding the MAC layer, the Physical Protocol Data Unit (PPDU) is pre-computed when the reset signal $rst1$ in Fig. 2 is asserted. Then, transmission starts when $rst1$ is deasserted and the transmit enable TX-EN in Fig. 2 is asserted. The disoverlapped packet generation and transmission limits the peak power consumption, and hence the (instantaneous) harvested power demand for extended range. As per the WiFi standard requirements, the packet is encoded via Direct Sequence Spread Spectrum (DSSS) by the Barker code spreading block in Fig. 2, and is transmitted at 1 Mbps. The PPDU computation does not add any latency to the communication since the 144-bit preamble is immediately available (hardwired or set by on-chip fuses) and masks the time taken by the sensing output encoding.

## III. Measurement Results

The proposed tag was designed in 180 nm with die photo in Fig.7a, and characterized with the testing setup in Fig. 7b on wafers covering corners (TT, SS, FF, SF and FS). As sown in Fig. 6a the occupied active area is 1.3 mm$^2$ and it is assembled via chip-on-board on a custom-designed PCB. The two SMDs, low-cost and commercially available RX-ANT (used for energy harvesting, clock extraction and position/motion detection operations) and the TX-ANT (for backscattering transmission) has a peak gain of 3 dBi with reflection coefficient, $S_{11,min}$ ~ -16 dB around the WiFi bandwidth. As for the adopted whip antenna for the PID and for the receiver, it shows a peak gain of 6 dBi and $S_{11,min}$ ~ -23 dB around the same application bandwidth. These last off-the-shelf antennas are typically used for WiFi routers and access points [3].

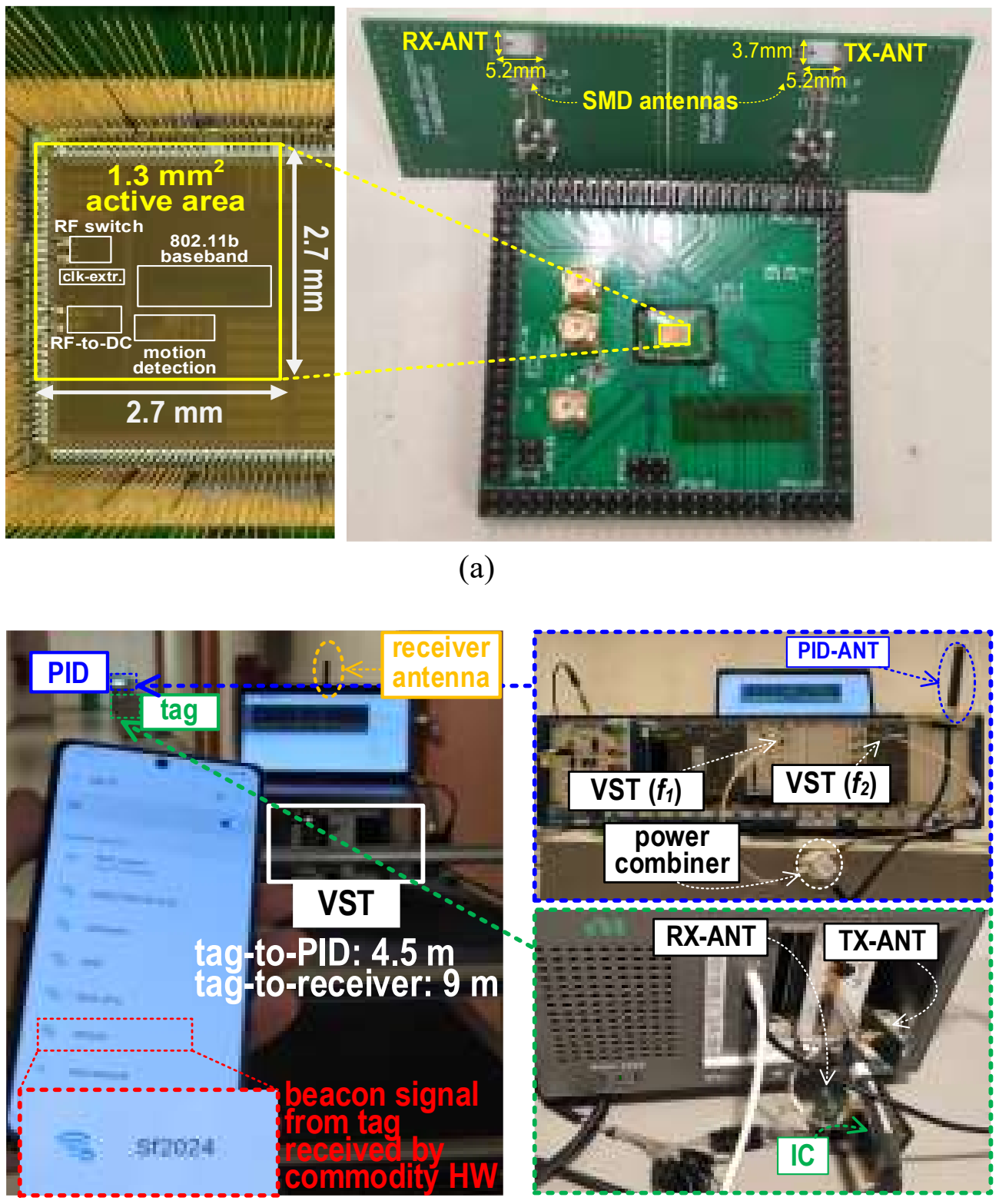

Fig. 7. a) Die micrograph and testing boards, b) testing setup.

### A. RF Harvesting and Position/Motion Detection

The measured harvested voltage dependence on the tag-to-PID distance is illustrated in Fig. 8a. The plot has a linear region with 110-mV/dBm slope between ~1 m and ~4 m, which corresponds to its operation under the slow-switching limit (SSL) condition[2] [39]. At shorter distances, $V_{harvest}$ saturates due to the ESD protection, which was designed to be activated above 2 V. At longer distances, $V_{harvest}$ becomes more sensitive to the distance due to operation under the fast switching limit[3] (FSL) [39].
From Fig. 8b, the harvesting sensitivity can be found from the minimum input (peak) power able to sustain uninterrupted system operation. In particular, an adequate DC voltage $V_{harvest}$ =0.66 V (i.e., above $V_{DDH}$) at a pessimistic system supply current $I_{LOAD}$=1.5 µA is fully provided at a two-tone input power down to -18.7 dBm. The latter corresponds to a maximum tag-to-PID distance of ~4.5 m under a PID total (two-tone) radiated power of 26-dBm, which is well within the FCC regulation limit.

[2] In the SSL condition, the charge pump output resistance is dominated by capacitive effects and remains constant due to the fixed operating frequency, making the output voltage linear with the input voltage.

[3] In the FSL condition, the charge pump output resistance is affected by the input voltage-dependent NMOS switch ON-resistance, leading to an overall super-linear dependence of the output on the input voltage

TABLE I. COMPARISON WITH PRIOR ART ON RF-TO-DC CONVERTERS OPERATING AT 2.4 GHZ (BEST OR NOTABLE PERFORMANCES IN BOLD)

| | This work | JSSC 2019 [19] | TCAS-I 2020 [20] | TMTT 2021 [22] | ACCESS 2023 [23] | TCAS-I 2023 [21] |
|---|---|---|---|---|---|---|
| technology (nm) | 180 | 65 | 180 | 180 | 65 | 65 |
| frequency (GHz) | 2.4 | 2.4 | 0.9, 2.4 | 2.4 | 1.9, 2.4 | 0.9, 1.9, 2.4 |
| topology | **crossed-NMOS Dickson** | cross-coupled differential drive | cross-coupled differential drive | Dickson | Dickson | cross-coupled differential drive |
| number of stages | 3 | 1 | 3 | 1 | 4 | 6 |
| sensitivity (dBm) | **-20.3 @ 1 V, 10 MΩ** | -17 @ 0.4 V, load N/A | -16 @ 1 V, 100 kΩ | -10 @ 1 V, load N/A | -13.2 @ 1 V, 500 kΩ | -19 @ 1 V, 800 kΩ |
| max PCE (%) | 43 @ **-10.7 dBm**, $R_L$=69 kΩ | 43.8 @ -3 dBm, $R_L$=1.6 kΩ | 47.1 @ -6 dBm, $R_L$=30 kΩ | **53.8** @ 0 dBm, $R_L$=30 kΩ | 32 @ 0 dBm, $R_L$=80 kΩ | 30.4 @ 4 dBm, $R_L$=100 kΩ |
| power efficiency FOM [(a)] | **890.1** | 570.3 | 753.6 | 538 | 422.4 | 577.6 |

[(a)] $FOM = PCE_{max}\frac{V_{OUT}\text{@sensitivity}}{P_{IN,LIN}\text{@sensitivity}}$

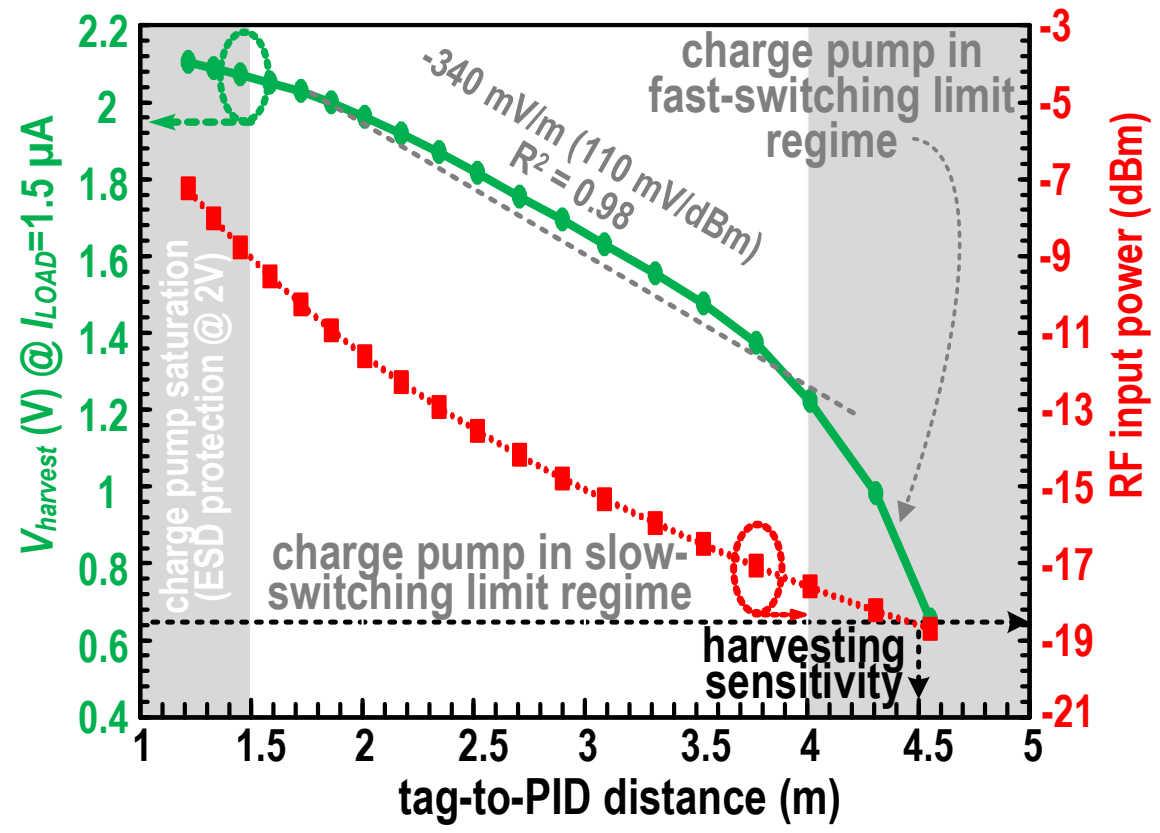

(a)

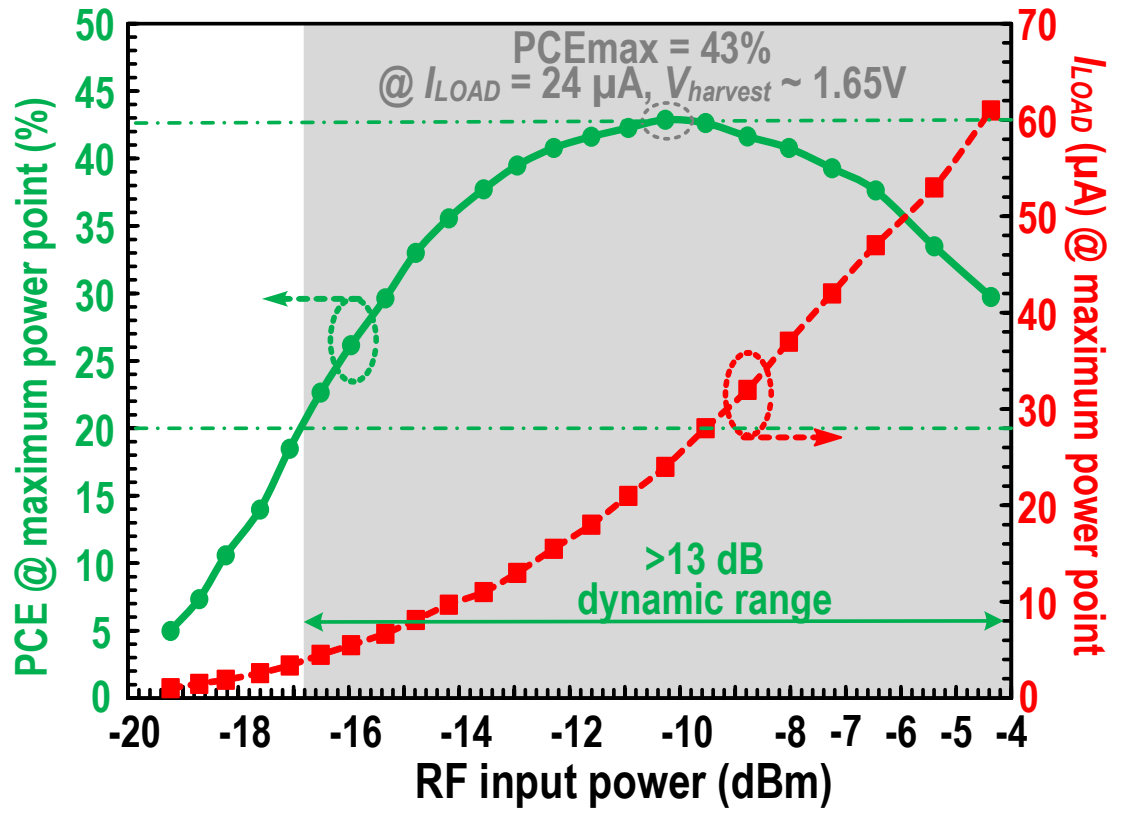

(b)

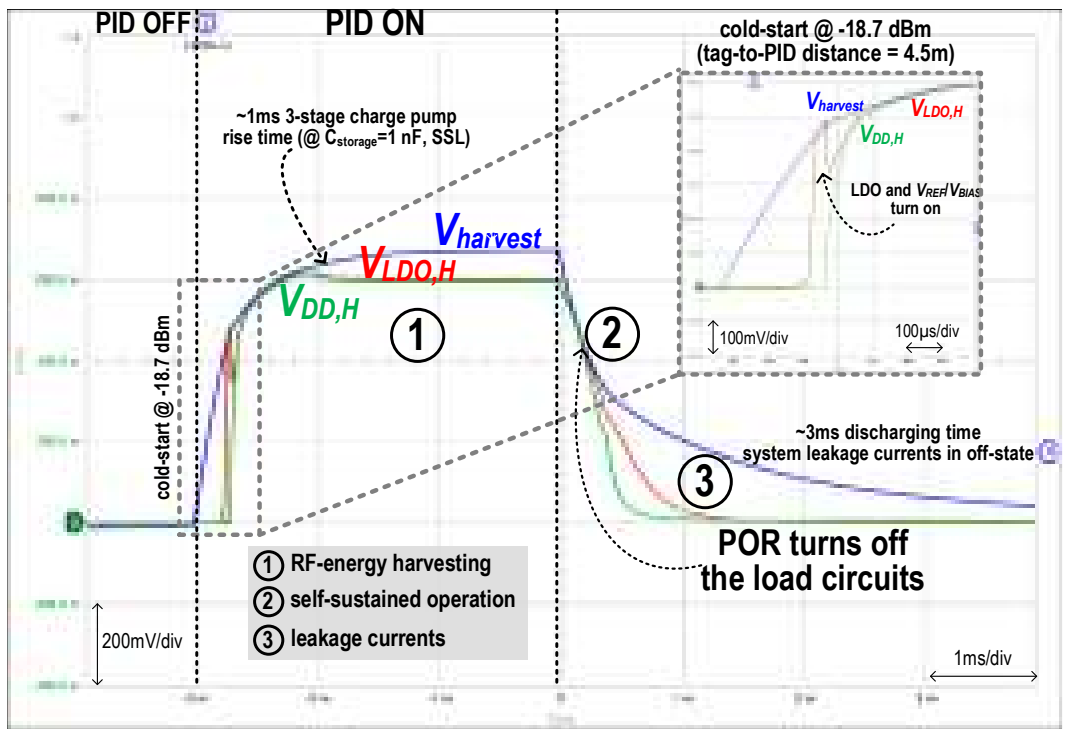

(c)

Fig. 8. Charge pump measurements: a) harvested voltage vs. tag-to-PID distance (26-dBm total power from PID, two-tones input 23-dBm/each), b) power conversion efficiency (PCE) at maximum power point vs. RF input power (two-tone input), c) waveforms of main power management signals (wake-up/shut-down cycles) including the cold-start condition @ -18.7 dBm (harvesting system sensitivity).

pump exhibits the best sensitivity Compared to prior ultra-low power RF-to-DC converters operating in the 2.4-GHz band in Table I, the proposed charge with an advantage of 1.3 dBm over the prior best [21]. At the same time, it achieves 43% peak efficiency at a much lower input power of -10.7 dBm, which is 4.7-14.7 dBm better than prior art in Table I (corresponding to 2-4X shorter distance from Fig. 8a). In other words, the proposed RF-to-DC conversion is able to sustain full system operation over a lower RF power, removing the spatial restrictions imposed by prior RF-to-DC converters. Overall, from Table I the proposed charge pump exhibits a 1.2-2.1X improvement in the power efficiency *FOM* (re-adapted from [40404040]) over prior art [19]-[23].

Regarding the transient response, the output rise time of the proposed charge pump is ~1 ms (0.5 ms) when operating in FSL (SSL). Hence, the rise time is confirmed to be well below the typical 10 ms-range latency target in WiFi environments.

Fig. 8c illustrates some key waveforms during the power-up stage when a new two-tone signal reaches the RX-ANT placed at a distance of 4.5 m from the PID, and the subsequent power-down. In phase 1, a new incident wave from the PID starts the harvesting process (including the cold-start condition), leading to the increase in $V_{harvest}$. If the latter raises above 0.48 V depending on the tag-to-PID distance (see Section IIA), the Power-On Reset is activated and triggers supply regulation and power provisioning to the entire system. After the tag has transmitted back, the PID is turned OFF (phase 2) and graceful power-down takes place by draining some of the energy left in the capacitor $C_{storage}$. Complete shut-down of all peripherals takes place when $V_{harvest}$ crosses the 0.42-V threshold (phase 3), when $C_{storage}$ is completely discharged by leakage in approximately 3 ms.

Regarding the position/motion detection sub-system, the linear part of the measured transfer characteristics of the circuit in Fig. 5 is plotted in Fig. 9a (the non-linear portion was shown in Fig. 8a). From this figure, the counter output has a linear trend with respect to the tag-to-PID distance, as confirmed by a coefficient of determination $R^2$ >0.997 across all process corners of their linear fit. The post-trimming discrepancy in the slope of the characteristics across corners is consistent and within 5% of the TT corner, which is 98 counts/m. From Fig. 9a, the worst-case temporal noise-induced uncertainty is 16 counts, which correspond to 16 cm in position detection.

Interestingly, such spatial resolution is competitive

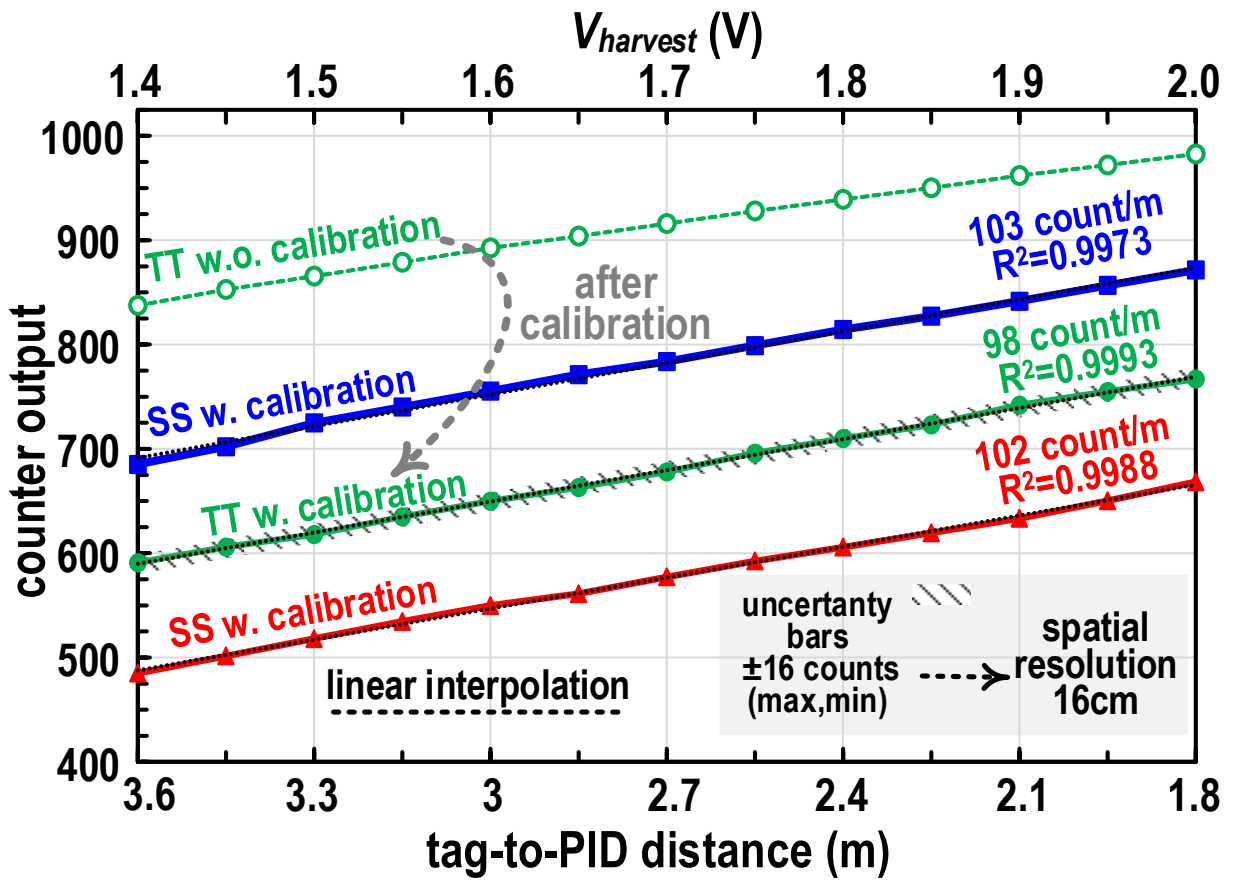

(a)

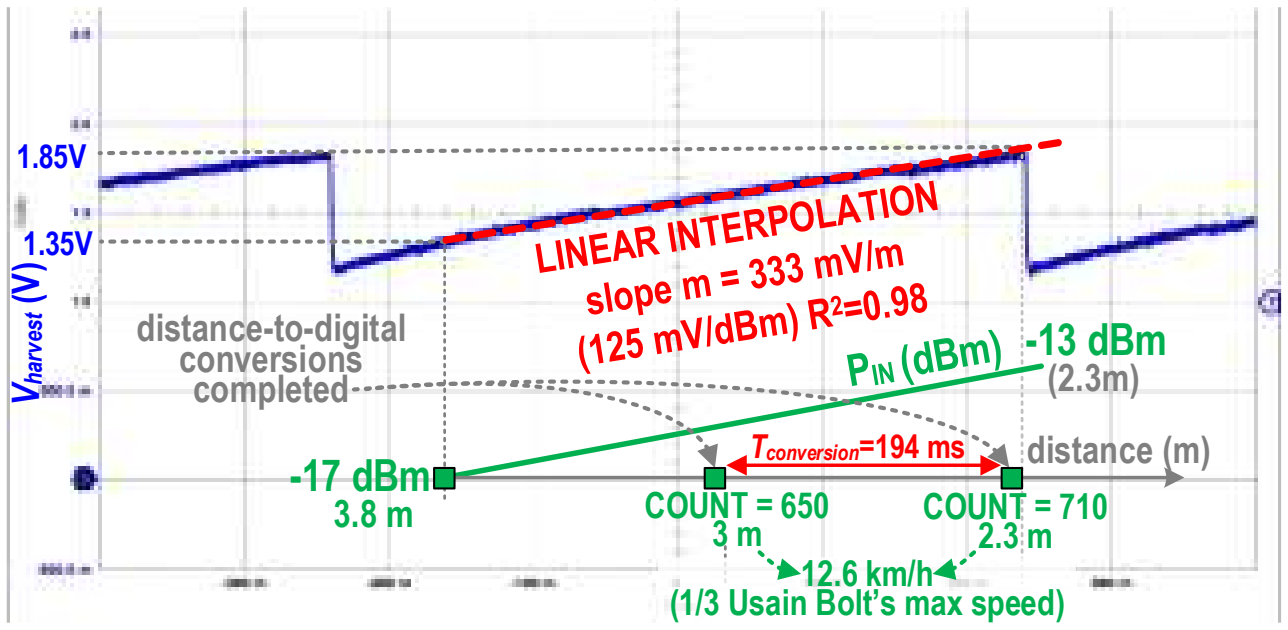

(b)

Fig. 9. Position/motion detection measurements: a) digital output vs. distance characteristics across extreme process corners (linear range in Fig. 8a), b) $V_{harvest}$ waveform in moving tag and resulting speed measurement.

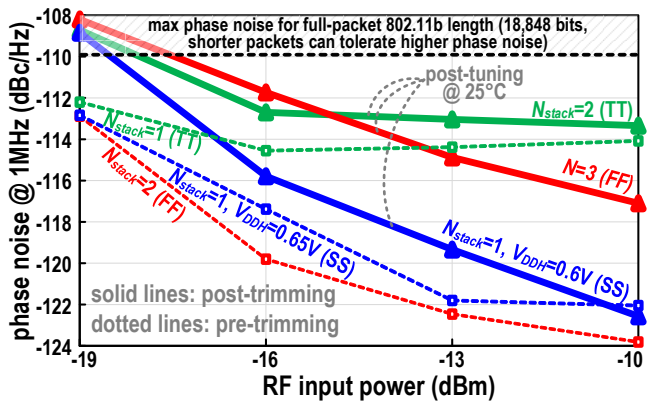

(a)

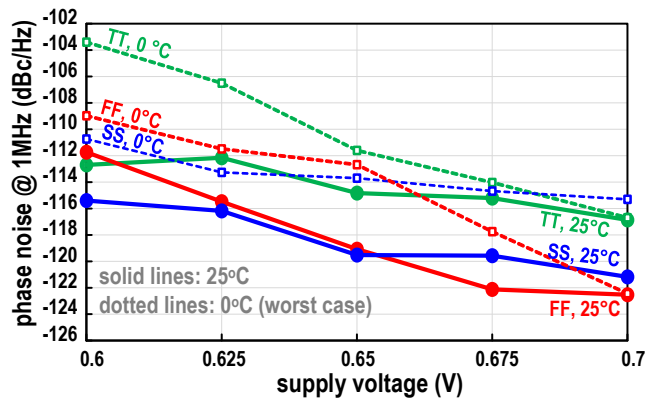

(d)

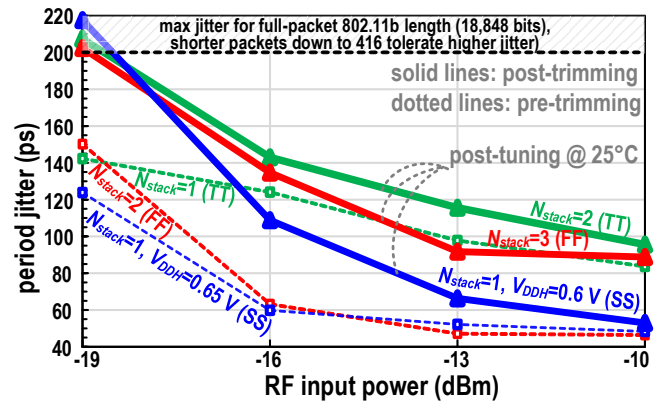

(b)

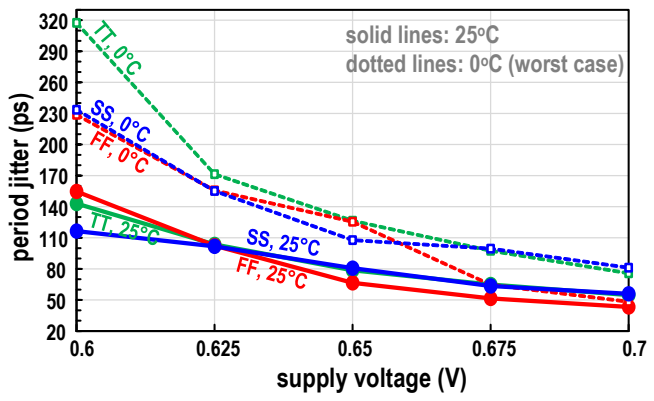

(e)

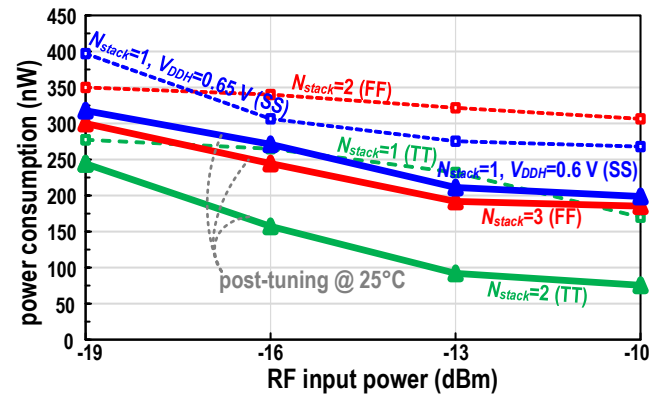

(c)

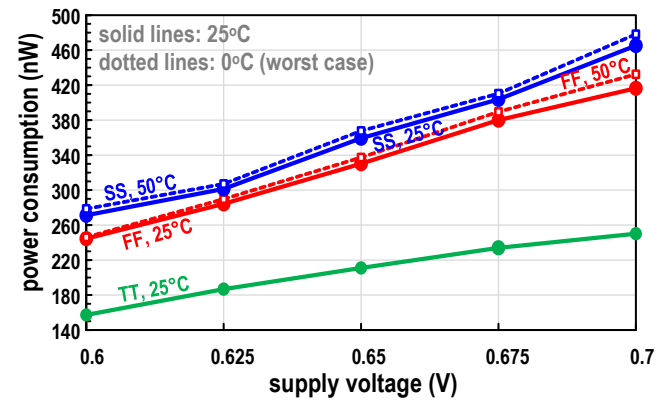

(f)

Fig. 10. Clock extractor measurements confirming 802.11b compliance: a) phase noise vs. input power (sum of two-tone signal power), b) jitter vs. input power (sum of two-tone signal power), c) power consumption vs. input power (sum of two-tone signal power) under different jitter-power settings. Resilience against voltage variations: d) phase noise, e) jitter and f) power consumption vs. supply voltage.

compared to typical meter-range resolution in conventional RSS-based position detection [26]. Speed detection of moving object shown in Fig. 9b, where speed can be measured up to 12.6 km/h, as limited by the <200-ms conversion time of the circuit in Fig. 5. The 1-nJ energy/conversion is a minor fraction (~0.5%) of the total system energy per sensing/transmission cycle (< 189.5 nJ). It is worth noting that the position/motion detection measurements are executed in x-axis alignment condition between the tag RX-ANT and the PID-ANT, because of the unidirectionality of the latter adopted SMD antenna.

In addition, the robustness of the position detection system is verified against interferences signal up to a tones-to-interference ratio equals to 10 dB, resulting in a worst-case inaccuracy of 15 cm (below the 16 cm-spatial resolution shown in Fig. 9b). In any case, this condition is not limiting at all, since the typical signal power (e.g., WiFi from access points or routers) is around -30 dBm, measured at distance of 0.5 m away from the interfering source and it is more than 10 dB below the proposed system harvesting/clock-extraction and position detection sensitivity (shown in Fig. 8a as -18.7 dBm at 4.5 m tag-to-PID distance).

As regard the accuracy of the motion (speed) detection, the worst-case and pessimistic condition is represented by an interference affecting just one of the two successive-in-time tag position detections. Indeed, if the just the first (second) position detection is affected by an interfering signal, resulting in an inaccuracy of 15 cm as stated above, the motion detection will have the same uncertainty level, giving hence an inaccuracy of 0.54 km/h (around 4.3% of the maximum detectable speed equals to 12.6 km/h as in Fig. 9b).

*B. Clock extractor/Two-tone Backscattering TX*

The measurement results in Fig. 10a-g show that the clock extractor complies with the specifications of the 802.11b WiFi standard. In particular, the phase noise and cycle-to-cycle jitter requirements are met across the entire input power range, and across all process corners from Figs. 10a-b. Their worst-case value is expectedly achieved at the minimum input power at which the system operates correctly as in Fig. 8a (-18.7 dBm), which corresponds to the maximum tag-to-PID distance within the same figure (4.5 m). Phase noise and cycle-to-cycle jitter obviously improve at shorter distances and hence larger input power levels, as shown in Figs. 10a-b.

From Fig. 10c, clock extraction power consumption ranges from 75 nW to 240 nW when the input power ranges from -19 to -10 dBm, and decreases at higher input power due to the increasing voltage input amplitude $V_{RF}$. Indeed, when $V_{RF}$ increases the clock extractor consumes lower (short-circuit) power, since its input voltage remains around the inverter logic threshold for a shorter time. Worst-case clock extraction power takes place in SS corner wafers, due to the larger trimmable supply voltage needed to sustain the targeted clock frequency (see Section II.B). Figs. 10a-c also show the ability to adjust the jitter-power tradeoff via supply voltage and $N_{stack}$ trimming, as discussed in Section II.B. For example, the un-necessary excess of phase noise and jitter performance over the WiFi requirement in Figs. 10a-b of SS corner at $N_{stack}$=1 and 0.65 V

TABLE II. COMPARISON WITH STATE-OF-THE-ART BACKSCATTERED TRANSMITTERS (BEST PERFORMANCE OR NOTABLE FEATURE IN BOLD)

| | | This Work | ISSCC'24 [29] | ESSCIRC'23 [11] | ISSCC'23 [14] | ISSCC'23 [13] | VLSI'21 [4] | ISSCC'20 [12] | JSSC'20 [10] | ISSCC'15 [9] |
|---|---|---|---|---|---|---|---|---|---|---|
| die | technology (nm) | 180 | 65 | 180 | 65 | 65 | 180 | 65 | 65 | 65 |
| | active area (mm$^2$) | 1.3 | 0.95 | 1.92 | 0.18* | 0.43 | 1.62 | 0.34 | 0.013 (dielet) | 0.26 |
| | evaluated under corner wafers (process variations) | **YES** | NO | NO | NO | NO | **YES** | NO | NO | NO |
| wireless | wireless protocol (RX/TX) | WiFi | WiFi (harvesting), WiFi (RX), BLE(TX) | BLE5 LE coded (S=8) | BLE | LTE (harvesting), BLE (RX), WiFi (TX) | WiFi | WiFi | custom | custom |
| | carrier frequency (GHz) | 2.4 | 2.4 | 2.4 | 2.4 | 2.4 | 2.4 | 2.4 | 5.8 | 5.8 |
| | type of backscattering | passive | passive | passive | passive | Hitchhike | passive | Hitchhike | Hitchhike | Hitchhike |
| | data-rate (Mbps) | 1 | 1 | 0.125 | 2 | 1 | 1 | 2 | 0.0042 | 2.5 |
| | TX range (m) | 9-25** | N/A (GFSK) | 97 (GFSK) | 160 | 0.5 | 8 | 90 | 0.001 | 0.1 |
| frequency synthesis/extraction | clock generation circuit | **clock extractor from 2-tone incident wave** | phase-flip tracking clock recovery with ILRO | self-sampling DCO with ring.osc. | external clock generator | SAR-FLL ring osc. | PVT-compensated DCO | PLL | clock extractor | XO |
| | local clock frequency (MHz) | 11 | 8 | ~24 | 4.25 | 11 | ~11 | 25 | 20 | 96 |
| | jitter stdev | **≤207 ps*** (compliant)** | N/A | ≤260 ps (compliant) | N/A | N/A | 220 ps (compliant) | N/A | N/A | N/A |
| | phase noise (dBc/Hz) | **-110 or better*** @ 1 MHz across corners** | ~ -93 @ 1 MHz (no corner) | N/A | N/A | N/A | -110 or better @ 1 MHz at TT corner | N/A | N/A | N/A |
| power management | # charge pump stages | 3 | N/A | N/A | 3 | 5 | N/A | N/A | 4 | N/A |
| | charge pump peak harvesting efficiency (%) | **43** | N/A | N/A | N/A | N/A | N/A | N/A | N/A | N/A |
| | RF harvesting sensitivity (dBm) | **-20.7** | N/A | N/A | N/A | N/A | N/A | N/A | N/A | -5.9 @ 0.6 V |
| system | avoidance of off-chip oscillator | **Y** | **Y** | **Y** | N (4 MHz) | N (32 kHz XO) | N (32 kHz XO) | N | **Y** | N |
| | on-chip sensors (on/off chip) | **position, speed, proximity (on-chip)** | N | temperature, capacitive, light (on-chip) | N | N | temperature, capacitive, light (on-chip) | N | N | motion, temperature (off-chip) |
| | RF signals needed simultaneously | tone (PID) | WiFi | tone (PID) | BLE (reversely-whitened, custom router) | LTE + BLE | tone (PID) | WiFi (XOR decode, custom router) | signal from custom coil reader | signal from custom coil reader |
| | tag peak power (μW) | **0.87 (1X)** | 17 (19.5X) | 10.6 (12.2X) | 7.16 (8.2X) | 25 (28.7X) | 2.5 (2.9X) | 28 (32.1X) | 10.6 (12.1X) | 113 (129.6X) |

* Estimated from die micrograph  ** Tag-to-PID distance = 1.5-4.5 m  *** Worst case across all distances within range and PVT corners

TABLE III. POWER/OPERATING DISTANCES SUMMARIZING TABLE

| | PID single-tone generated power (overall, two-tone) (dBm) | | | | |
|---|---|---|---|---|---|
| | 11 (14) | 14 (17) | 17 (20) | 20 (23) | 23 (26) |
| **max. tag-to-PID distance (m) (@-18.7 dBm)[a]** | 1.3 | 1.8 | 2.5 | 3.5 | 4.5 |
| **max. tag-to-receiver distance (m) (@ max. tag-to-PID distance)[b]** | 10 | 9.8 | 9.7 | 9.5 | 9 |
| **min. tag-to-PID distance (m) (@-18.7 dBm)[c]** | 0.45 | 0.65 | 0.9 | 1.25 | 1.5 |
| **max. tag-to-receiver distance (m) (@ min. tag-to-PID distance)[b]** | 30 | 29 | 27 | 26 | 25 |

[a] -18.7 dBm indicated as harvesting system sensitivity for interrupted operations (clock extraction, position/motion detection, WiFi backscattering communication).
[b] target BER=$10^{-5}$ (SNR=11 dB) at the receiver end for WiFi standard compliance.
[c] -10 dBm indicated as maximum tested power for the position/motion sensing (charge pump saturation at 2 V).

(FF, $N_{stack}$=2) can be sacrificed to reduce power by reducing the supply voltage to 0.6 V (increase $N_{stack}$ to 3), saving up to 27% (42%) power from Fig. 10c at -10 dBm input power and across corners. The same considerations are extended under voltage variations in Figs. 10d-f, confirming the compliance with the 802.11b requirements across voltages (0.6-0.7 V). From the same figures, compliance is also confirmed across temperatures (0-50°C, well beyond indoor conditions) for the longest allowed WiFi packet (18,848 bits). It is worth noting that further power can be saved under shorter packets, thanks to the more relaxed cycle-to-cycle jitter requirement (since less jitter is accumulated throughout shorter packets).

In addition, the robustness of the clock extraction is verified under a (two) tone-to-interference ratio equals to 10 dB as a worst-case and quite pessimistic scenario as stated in Sec. IVA for the position detection. The percentage variation of the period jitter of the extracted clock is, then, less than 5% when the RF input power for each tone is around -16 dBm.

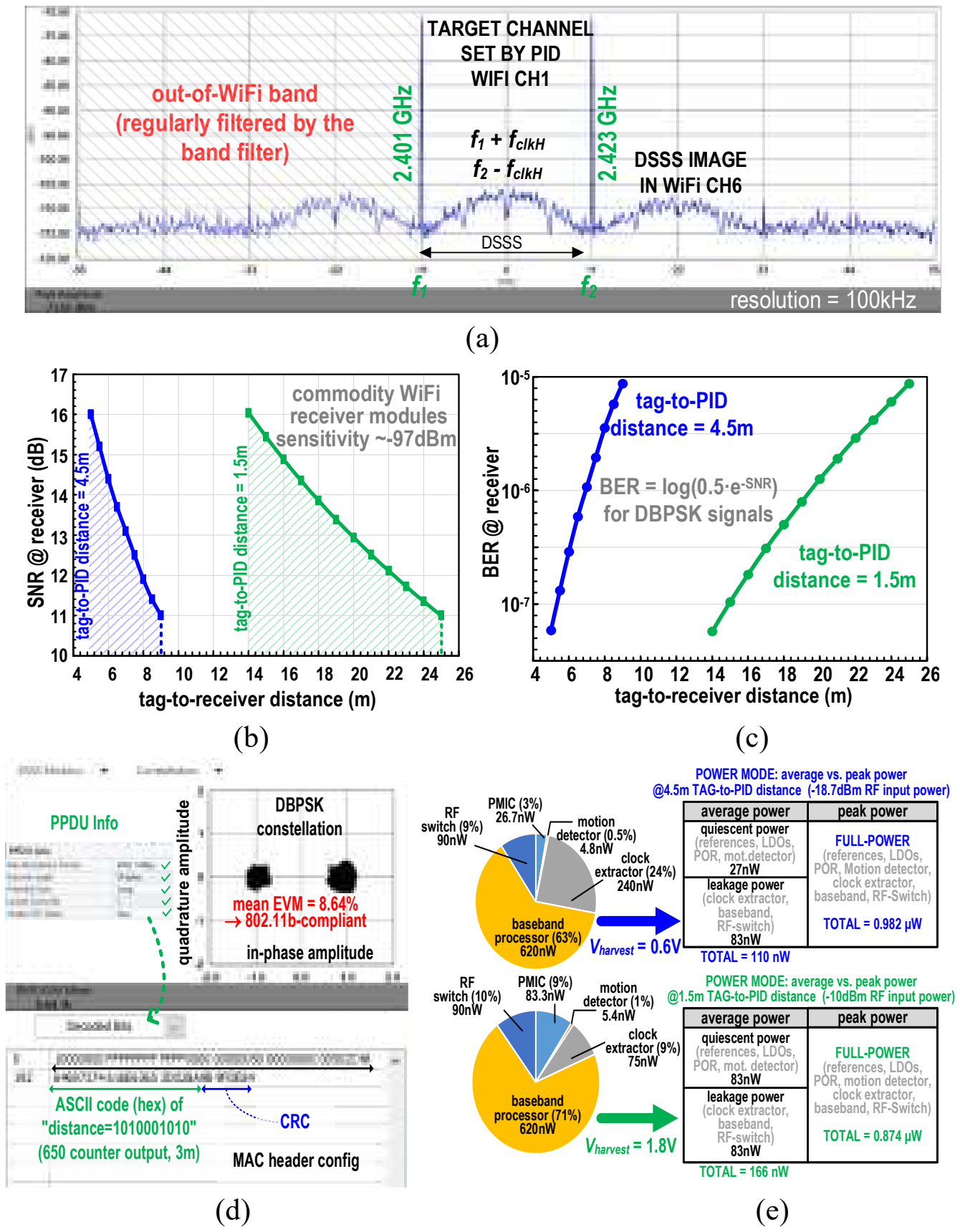

Fig. 11. Measurements: a) received DSSS spectrum in WiFi channel 1, b) $SNR$ vs. tag-to-receiver distance, c) resulting $BER$, d) DBPSK constellation, EVM and correct distance-related 39-byte packet decoding, e) system power breakdown at different tag-to-PID distance.

With regards to the backscattered transmission, WiFi compliance is shown in the received DSSS spectrum around the non-overlapped channel #1 of the 802.11b standard in Fig. 11a, while operating at 4.5-m tag-to-PID distance and 9-m tag-to-receiver distance. The received signal power is -86 dBm, and hence well within the sensitivity of commodity 802.11b receivers available commercially, which is typically significantly better (e.g., -97 dBm in [41], [42]). For completeness[4], the resulting $SNR$ is plotted versus the tag-to-receiver distance in Fig. 11b at a tag-to-PID distance of 4.5 m and 1.5 m (i.e., the extreme points of the interval considered in Fig. 8c). The corresponding pre-error correcting code $BER$ is evaluated as per the usual relation for the DBPSK modulation [43], [44]. From Fig. 11a the $BER$ is always better than $10^{-5}$, and hence complies with the 802.11b standard within the above considered range. In addition, the backscattering TX does not interfere with simultaneous the clock extraction operation since the transmitted spectrum is usually 25-30 dB below the peak power of the backscattered tones as showed in Fig. 11a and Fig. 6c.

The two constellation points of the DBPSK modulation in Fig. 11d are clearly distinguishable. More quantitatively, the measured error vector magnitude (EVM) of 8.64% is well within the 35% prescribed by the standard. The correct 39-byte payload reception under DSSS encoding at 1Mbps is shown in the packet report from the WiFi testing equipment in the same figure. For completeness, correct packet reception and decoding is shown in Fig. 10d while the measured tag-to-PID position (3 m) is transmitted through the RF-switch and detected by a receiver at 9m distance.

Finally, Tab. III summarizes the maximum and minimum distance ranges for the tag-to-receiver distance while varying both the tag-to-PID distance and the PID power as well.

From Fig. 11e, clock extraction at 2-V (0.6-V) harvested voltage and thus near (far) the PID consumes only 8.5% (25%) of the overall power, as opposed to 20% (37% including the necessary voltage regulation loop) in prior WiFi backscattering transmitters with lowest power [4]. System power is dominated by dynamic power, being leakage power always below 20%.

Comparison with the state of the art in Table II shows that the proposed architecture uniquely reuses the incident RF wave for multiple purposes, i.e. harvesting, clock extraction, position/motion detection and backscattering transmission. The proposed architecture and the suppression of the on-chip oscillator via two-tone clock extraction reduces the peak power to 0.87 μW, which is 2.9-129X lower than prior art. In spite of the lower power, the extracted clock exhibits better jitter performance than [11] by 53 ps, and [4] by 23 ps. These results, as in [4], uncommonly include the impact of process corner wafers, which was disregarded in all prior art.

[4] Most prior art does not report $SNR$ versus distance [3], [4], [11]-[14], [27], [28], similarly for $BER$ (with the exception of [12] and [3]).

## IV. Conclusion

In this work, the first sub-μW 802.11b backscattering transmitter has been presented and experimentally validated under all process corner wafers (corners not considered in prior art). Its architecture reuses the same incident wave for RF harvesting, backscattering communications, clock extraction and position/motion sensing. Such reuse removes the battery, any explicit physical harvester, any power-hungry on-chip local oscillator, and off-chip motion sensor (e.g., MEMS) for aggressive miniaturization, unrestricted device lifespan, low cost and low maintenance cost for ubiquitous adoption. Finally, the adopted clock-extraction and two-tone backscattering communication open the road to the adaptability of the proposed battery-less tag to other standards such as BLE, LTE and 5G.

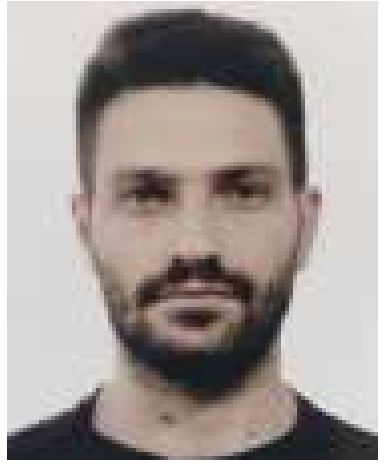

**Marco Privitera** (Member, IEEE) was born in Piazza Armerina, Italy, in 1997. He received the B.S. and M.Sc. degree (summa cum laude) in electronic engineering from the University of Catania, Catania, Italy in 2019 and 2021, respectively, where is currently pursuing the Ph.D. degree. In 2023 he joined the Green-IC research group within the National University of Singapore (NUS), Singapore, as Visiting Ph.D. student.

His current research interests include power management integrated circuits, data communication sub-systems, and ultra-low power and ultra-low voltage analog blocks for energy harvesting applications. In 2023, he received the IEEE-CAS Pre-Doctoral Award.

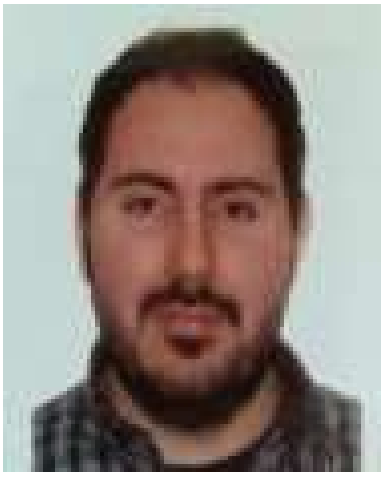

**Andrea Ballo** (Member, IEEE) was born in Catania, Italy, in 1990. He received the Laurea (summa cum laude) and Ph.D. degrees in electronic engineering from the University of Catania, in 2016 and 2020, respectively. Since 2021, he has been a Research Fellow and an Adjunct Professor of electronic devices with the University of Catania.

His current research interests include low-voltage low-power analog circuit design and analog and mixed electronics for energy harvesting applications. He is an Associate Editor of the *Journal of Circuits, Systems, and Computers*; a member of the editorial board of *Journal of Electronics and Electrical Engineering* (Universal Wiser Publisher); and a member of the topical advisory panel of different journals, such as *Applied Sciences* (MDPI) and *Electronics* (MDPI).

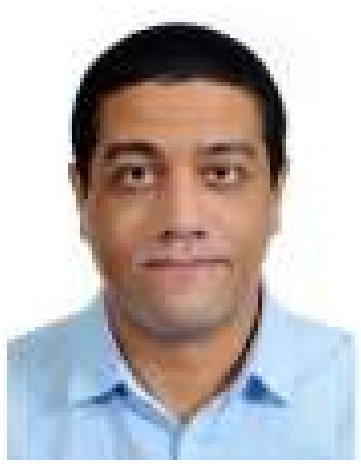

**Karim Ali** (Member, IEEE) received his bachelor's degree in electronics and communication engineering from the Fayoum University, Fayoum, Egypt, in 2010, and M.Sc. degree in electronics engineering from American university in Cairo, Cairo, Egypt in 2013 and Ph.D. degree in electrical and Computer engineering from National University in Singapore (NUS), Singapore, in 2019.

He is currently a senior research fellow at electrical and computer engineering department, National University of Singapore. He held various positions previously at Intel corporation, Oregon, USA, and Mentor Graphics, Cairo, Egypt, and Middle East Technical university, Cyprus. He has authored or co-authored more than 30 publications on Journals and conferences proceedings. He is an author of a book chapter and three patents. His research interest includes self-powered sensor nodes, non-volatile memories and ultra-low power analog circuits

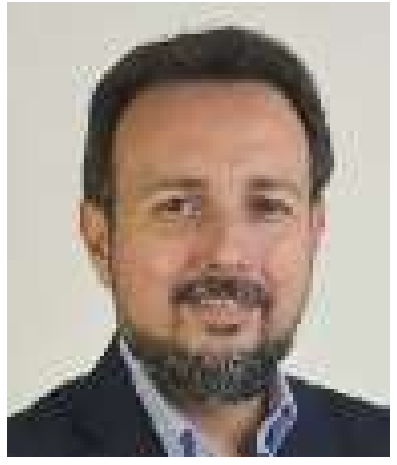

**Alfio Dario Grasso** (Senior Member, IEEE) was born in Catania, Italy, in 1978. He received the Laurea (summa cum laude) and Ph.D. degrees in electronic engineering from the University of Catania, Catania, Italy, in 2003 and 2006, respectively.

From 2006 to 2011, he worked as a Freelance Engineer in the field of electronic systems. From 2009 to 2010, he was an Adjunct Professor of electronics with the University Kore of Enna, Italy. He was appointed as Assistant Professor and Associate Professor in 2011 and 2015, respectively, and he is now Full Professor at the University of Catania. He teaches graduate courses on advanced VLSI digital design, microelectronics, and basic electronics. He has coauthored more than 130 papers on referred international journals and conference proceedings. His current research interests include low-voltage low-power analog circuit design and analog and mixed signal processing for energy harvesting applications.

Dr. Grasso is an Associate Editor for the *IET Electronics Letters*, Elsevier *International Journal of Circuits Theory and Applications*, and is a Member of the editorial board of MDPI *Sensors*.

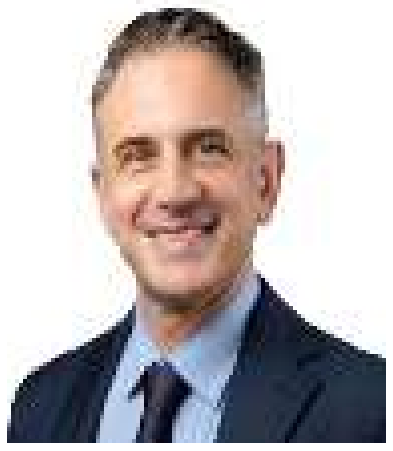

**Massimo Alioto** (M'01–SM'07-F'16) received the MSc degree in Electronics Engineering and the Ph.D. degree in Electrical Engineering from the University of Catania (Italy) in 1997 and 2001. He is currently Provost's Chair Professor at the Department of Electrical and Computer Engineering, National University of Singapore, where he leads the Green IC group, and is the Director of the Integrated Circuits and Embedded Systems area. Previously, he held positions at the University of Siena, Intel Labs – CRL (2013), University of Michigan Ann Arbor (2011-2012), BWRC – University of California, Berkeley (2009-2011), and EPFL (Switzerland, 2007).

He has authored or co-authored more than 380 publications on journals, conference proceedings, and five books, including the popular *Enabling the Internet of Things - from Circuits to Systems* (Springer, 2017). His primary research interests include widely energy-scalable systems, sustainable & self-powered silicon systems, data-driven systems for machine intelligence and hardware security, among the others.

He was the Editor in Chief of the IEEE TRANSACTIONS ON VLSI SYSTEMS (2019-2022), and the Deputy Editor in Chief of the IEEE JOURNAL ON EMERGING AND SELECTED TOPICS IN CIRCUITS AND SYSTEMS (2018). He is the Chair of the Distinguished Lecturer Program for the IEEE Circuits and Systems Society (2023-2024), and was Distinguished Lecturer for the same Society (2022-2023, 2009-2010) and the Solid-State Circuits Society (2020-2021). He was also member of the Board of Governors of the IEEE Circuits and Systems Society (2015-2020), and Chair of the "VLSI Systems and Applications" Technical Committee (2010-2012). In the last five years, he has given 50+ invited talks at top conferences, universities and leading semiconductor companies. His research has received various best paper awards (e.g., ISSCC, ICECS), public recognition from industry (e.g., among 10 technology highlights in TSMC annual report in 2020), and top contributor at the VLSI Symposium on Circuits (2023). He served as Guest Editor of several IEEE journal special issues (e.g., JSSC, TCAS-I, TCAS-II, JETCAS). He also serves or has served as Associate Editor of a number of IEEE and ACM journals. He is/was Technical Program Chair (e.g., ISCAS, SOCC, ICECS, NEWCAS, APCCAS), Track Chair in a number of conferences (ICCD, ISCAS, ICECS, VLSI-SoC, APCCAS, ICM), ISSCC and ASSCC subcommittee member. Prof. Alioto is an IEEE Fellow.